# CSLNSpeech: solving extended speech separation problem with the help of Chinese sign language

*Jiasong Wu[1,2,3,4,*], Xuan Li[1,2,4], Taotao Li[1,2,4], Fanman Meng[1,2,4], Youyong Kong[1,2,4], Guanyu Yang[1,2,4], Lotfi Senhadji[2,3,4], Huazhong Shu[1,2,4]*

[1]*Laboratory of Image Science and Technology, Key Laboratory of New Generation Artificial Intelligence Technology and Its Interdisciplinary Applications (Southeast University), Ministry of Education, Nanjing 210096, China*
[2]*Jiangsu Provincial Joint International Research Laboratory of Medical Information Processing, Southeast University, Nanjing 210096, China*
[3]*Univ Rennes, INSERM, LTSI - UMR 1099, F-35000 Rennes, France*
[4]*Centre de Recherche en Information Biomédicale Sino-français (CRIBs), Univ Rennes, INSERM, F-35000 Rennes, France*

ABSTRACT

Previous audio-visual speech separation methods use the synchronization of the speaker's facial movement and speech in the video to supervise the speech separation in a self-supervised way. In this paper, we propose a model to solve the speech separation problem assisted by both face and sign language, which we call the extended speech separation problem. We design a general deep learning network for learning the combination of three modalities, audio, face, and sign language information, for better solving the speech separation problem. To train the model, we introduce a large-scale dataset named the Chinese Sign Language News Speech (CSLNSpeech) dataset, in which three modalities of audio, face, and sign language coexist. Experiment results show that the proposed model has better performance and robustness than the usual audio-visual system. Besides, sign language modality can also be used alone to supervise speech separation tasks, and the introduction of sign language is helpful for hearing-impaired people to learn and communicate. Last, our model is a general speech separation framework and can achieve very competitive separation performance on two open-source audio-visual datasets. The code is available at https://github.com/iveveive/SLNSpeech.

## 1. Introduction

### *1.1. Speech separation*

The speech separation problem or cocktail party problem (Cherry, 1953) refers to an agent or a robot that focuses on a specific speech in a noisy environment by separating different categories of sounds from signals containing multiple voices and noises. The speech separation task is widely used for speech recognition (Deng et al., 2013; Fahad et al., 2020; Saudi et al., 2019), enhancement (Loizou, 2013; Li et al., 2020; Islam et al., 2020), dereverberation (Naylor and Gaubitchet, 2010). In speech recognition and speech enhancement systems, multi-speaker speech separation (Liu et al., 2013; Khan and Milneret, 2013; Wu et al., 2019) is considered to be the most challenging task because of the high correlation of the time-frequency structure between target and interfering voices, which means the acoustic features of target speech are easily confused with those of noisy speech (Reddy and Raj, 2007). Another reason that contributes to increasing this challenge is the high non-stationarity of multi-speaker speech, which leads to variations in noise estimation and higher complexity for predicting the target speech. This task becomes more challenging when there is only one channel for the mixed voice because the audio signal lacks information about the source location of the voice (Gu, 2010). Therefore, this paper focuses on the multi-speaker monaural speech separation, in which the interference source is other voices and noise.

Fig. 1 (a) shows a simple example of the classical speech separation process of two speakers. In this case, the agent or robot only receives audio information, called the audio-only speech separation (AOSS) method. Many recent works explore the use of audio-visual speech separation (AVSS) algorithms (Fig. 1 (b)), which introduce visual information to guide speech separation. The natural synchronization between vision and sound can provide rich supervision signals for locating sound in vision (Aytar et al., 2016). In the multi-speaker speech separation problem, facial features and lip movement information are common visual supervision features. The inspiration for audio-visual speech separation comes from the fact that people pay attention to active speakers with the help of visual information in noisy environments. We have noticed that the movement of gestures can also attract people's attention. Thus, we introduce sign language, a standard gesture definition, into speech separation. As shown in Fig. 1 (c), we propose a general model named audio-visual-sign speech separation (AVSSS), which can simultaneously receive the audio, face, and sign language features as input in noisy environments and separate the voices of different speakers. It is worth noting that our model can still separate speech when there is no vision or sign language modality. In this paper, the word visual or vision generally refers to facial visual information, but this does not mean that we do not regard sign language information as a visual modality; it is just to be consistent with the expression of the audio-visual system in the previous works.

* *Corresponding author.* Tel.: 86-25-83794249; fax: 86-25-83792698.
E-mail address: jswu@seu.edu.cn

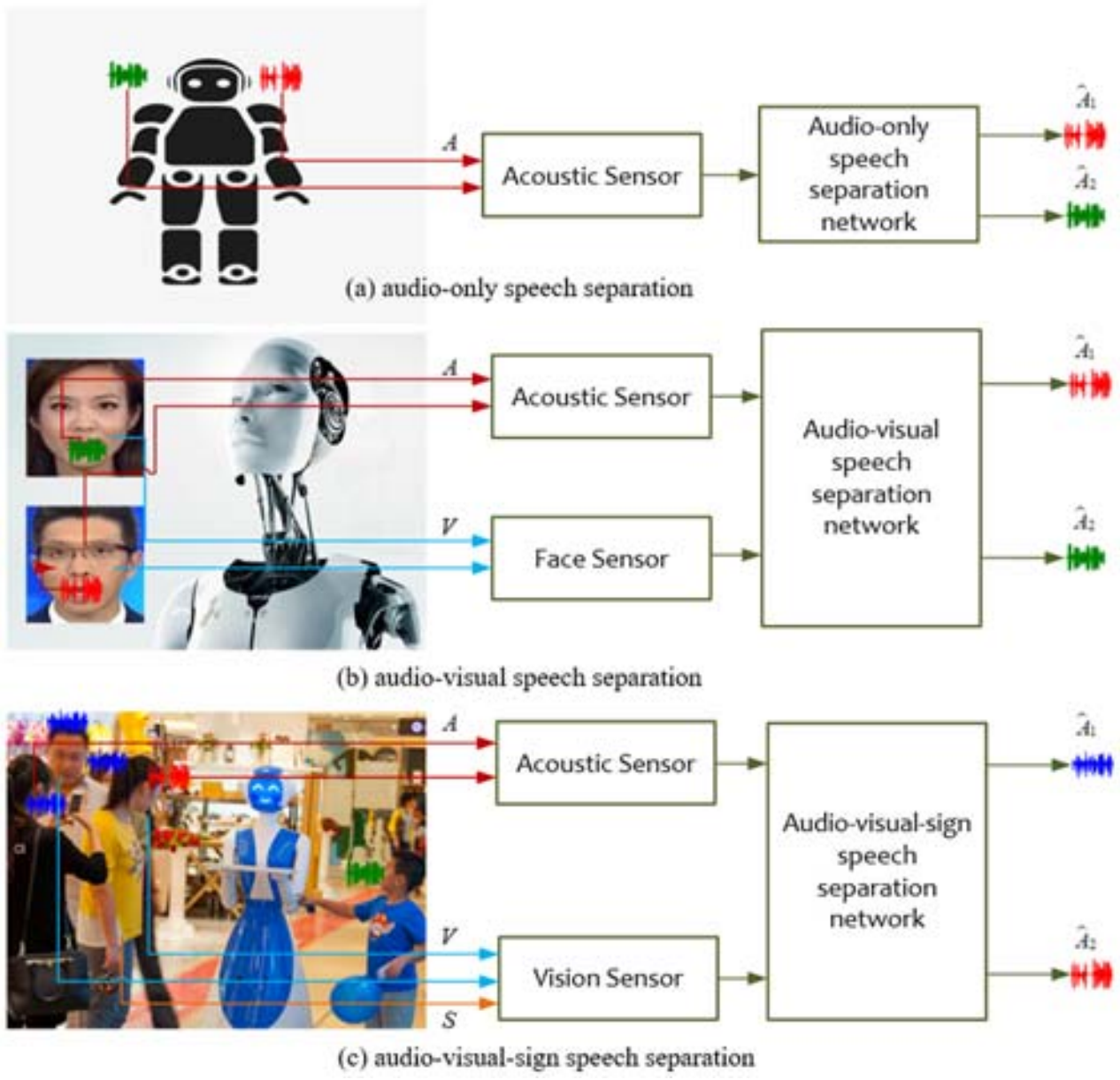

Fig 1 The comparison of audio-only, audio-visual, and audio-visual-sign speech separation. The red arrow tracks the audio information, the blue arrow tracks the face information, and the orange arrow tracks the sign language information. The data of multiple modalities is obtained through sensors (such as cameras), and then the voice of different speakers in the mixed audio is separated through the speech separation network.

### *1.2. The importance of sign language*

Sign language is a language that forms particular meanings or words according to the changes in gestures. It is one of the common ways of communication for people with hearing loss. People with hearing loss cannot hear the correct voice to varying degrees. Thus, their language learning ability is underdeveloped (Socher et al., 2019), resulting in many difficulties in social communication, such as medical scenarios (Barnett, 2002). Helping people with hearing loss develop speech skills can help solve their problems in social communication. Holt et al. (2023) show that audio-visual cues can improve the ability of children with hearing loss to process language. Connecting sign language with audio can also help people with hearing loss improve their language ability because sign language is as closely related to voice as lip movement. Thus, we introduce sign language to the speech separation task. This technology can remove background voices in noisy speech-teaching environments for people with hearing loss or make intelligent robots interact with hearing loss people with speech and sign language abilities.

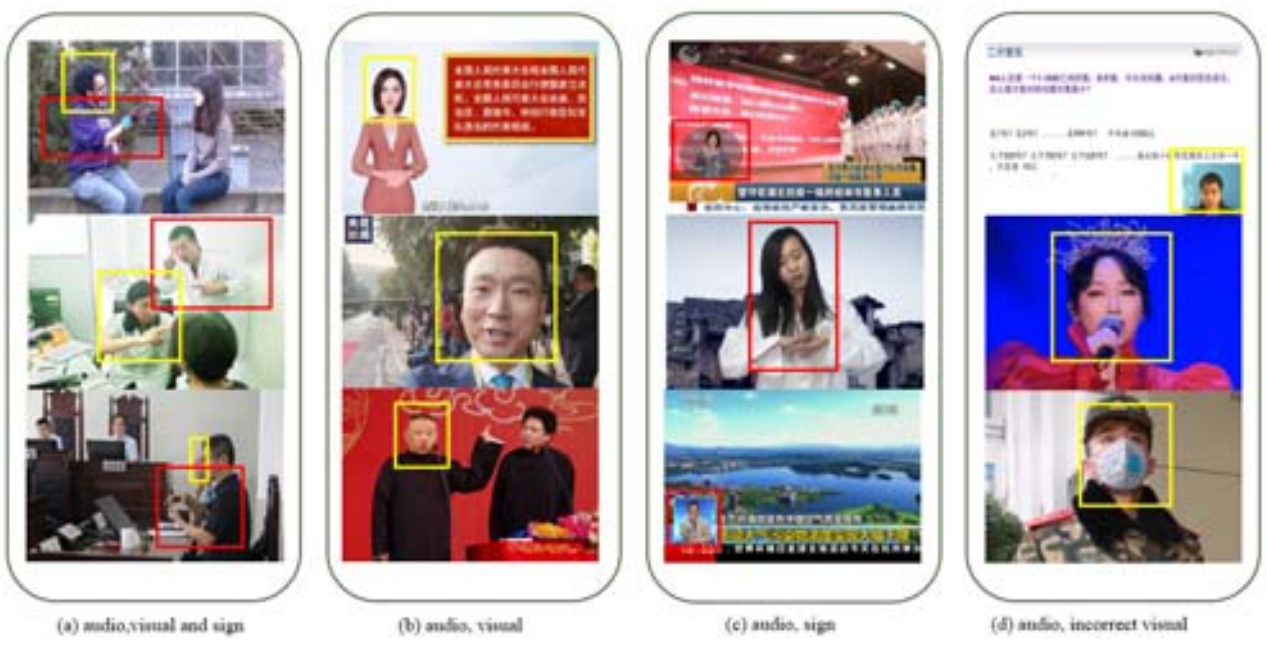

Fig 2 The coexistence of audio, sign language, and face in real-life scenarios where the yellow boxes indicate the face and the red boxes represent the sign language. (a) The scene of audio, face, and sign language modalities; (b) The scene of audio and face modalities; (c) The scene of audio and sign language modalities. (d) The scene of audio and the concealed face modalities.

Multiple modalities exist in many natural scenes. As reported in Fig. 2 (a), some coexisting scenes of three modalities, including the gloves for sign language translation, free clinics, and court of law for people with hearing loss to communicate with ordinary people. Fig. 2 (b) shows situations where audio and face exist simultaneously. The scenes where only the speaker's voice and the signer's sign language are collected can be seen in Fig. 2 (c). In these cases, we can collect time-aligned voice, face, and sign language videos that convey the same semantic information, which is the data basis of using multiple modalities to guide speech separation. However, these data are not always suitable. As in Fig. 2 (d), the soundtrack is not in sync with the picture in an online classroom. It is usual for objects such as microphones or masks to block the face, leading to the loss of face and voice synchronization and failure of the audio-visual speech separation system. Thus, we introduce sign language to improve the robustness of the system.

### *1.3. Towards sign language assisted speech separation*

As the first study to introduce sign language into speech separation, we call it the extended speech separation problem, and many challenges need to be addressed. We first focus on how sign language guides speech separation. Sign language can convey semantic information that is consistent with speech through actions, but it may not necessarily have the same word order. For example, negative words are placed at the end of sentences in Chinese sign language. Secondly, normal individuals and deaf-mute individuals may need the help of sign language translators when communicating, which means that the sign language and speech modes do not come from the same person. These characteristics are different from previous audio-visual speech separation systems. Therefore, in the extended speech separation problem, we refer to face supervision as strong supervision and sign language supervision as weak supervision. Secondly, we investigated the joint supervision of face and sign language on speech separation. This is meaningful because we have innovatively introduced sign language modalities and proposed a universal framework for learning and integrating these three modalities, which has better learning ability and robustness.

To be clear, we raise three questions as follows:

Table 1. Abbreviations in the paper

| Symbols | Brief descriptions |
|---|---|
| AO | audio-only |
| AOSS | audio-only speech separation |
| AS | audio-sign |
| ASSS | audio-sign speech separation |
| AV | audio-visual |
| AVS | audio-visual-sign |
| AVSS | audio-visual speech separation |
| AVSSS | audio-visual-sign speech separation |
| CASA | computational auditory scene analysis |
| CNN | convolutional neural network |
| DNN | deep neural network |
| FCN | fully convolutional network |
| MTCNN | multi-task convolutional neural networks |
| NMF | nonnegative matrix factorization |
| PCC | Pearson correlation coefficient |
| PESQ | Perceptual Evaluation of Speech Quality |
| RNN | recurrent neural networks |
| SAR | Sources to Artifacts Ratio |
| SDR | Source to Distortion Ratio |
| SIR | Source to Interactions Ratio |
| CSLNSpeech | Chinese Sign Language News Speech |
| SLR | sign language recognition |
| SS | speech separation |
| STFT | Short Time Fourier Transform |
| STOI | Short Time Objective Intelligibility |
| WLASL | Word-Level American Sign Language |

(1) Can sign language supervise speech separation like visual information does?

(2) Is the performance of AVSSS better than AVSS?

(3) Can the weakly supervised sign language improve the robustness of the AVSSS?

To address the above three issues, we propose an effective self-supervised learning framework to address the extended speech separation problem. Previous studies mainly used two approaches: using synchronous cues, most commonly the lip movements (Ephrat et al., 2018) of the target speaker, or using static cues, such as the face characteristics (Chung et al., 2020). The former uses dynamic evidence highly correlated with the expected speech to guide speech separation. However, when lip movement becomes unreliable, such as the mouth area being blocked by the microphone or the speaker turning their head away, speech separation relying solely on lip movement may fail. The latter is not affected by interruptions, but the discrimination may decrease, such as difficulty distinguishing two people with similar appearances. There are also studies (Gao et al., 2021) that combine lip movement and static facial features to supervise speech separation, which can slightly resist performance degradation when lip movement is unreliable. Considering the asynchronous nature of the word order in sign language, we choose not to use a complete video stream but to evenly select several frames of images at the same time interval. The time interval is referred to as time granularity and is controlled by hyperparameters. In this way, we can control how long a sign language frame should express the semantics of a speech segment. Meanwhile, we propose a multi-modal feature fusion method using spatial attention. In [3], the author extracts the facial embedding using resnet18 and replicates the facial embedding along the time dimension to match the speech features. Many studies have adopted this method of processing facial images. However, compressing image features into one-dimensional embeddings may lose information such as shape or texture that can be used to distinguish different people, significantly impacting speech separation. This article uses a spatial attention mechanism that largely preserves image features.

In summary, the contributions of the paper are as follows:

(1) A new multi-modality dataset, Chinese Sign Language News Speech (CSLNSpeech), is proposed. CSLNSpeech contains three time-synchronized modalities: audio, visual, and Chinese sign language. To the best of our knowledge, the CSLNSpeech dataset is the first dataset that gathers such synchronized modalities and thus opens the way to explore the joint and marginal characteristics of these data using self-supervised learning methods.

(2) Based on the CSLNSpeech dataset, we provide a standard benchmark for extended speech separation. The separation of mixed audio uses audio sources, Chinese sign language embedding, and visual embedding simultaneously. It is worth noting that in the absence of face or sign language modality, our model can still operate normally, and our audio-visual separation performance has produced the most advanced results on the two open-source audio-visual speech datasets.

(3) We explore the influence of Chinese sign language modality in speech separation for the first time and discuss how this modality can help improve separation performance and robustness. Different from the sign language recognition task, which is another highly challenging task, we only obtain the feature representation of sign language synchronized with speech and do not need to know the specific meaning of sign language, just like we do not need to know the specific content of speech in speech separation.

The rest of the paper is organized as follows. In Section 2, we review the speech separation methods and related datasets. In Section 3, we construct the new dataset we named CSLNSpeech. Section 4 presents our model that leverages sign language and visual information as a supervisory signal for speech separation. Section 5 presents several experiments to analyze our model and provide the benchmark of CSLNSpeech. Section 6 provides a discussion of the research. The conclusion is formulated in Section 7. Table 1 gives a list of abbreviations used in this paper and a brief description.

## 2. Related work

Research that has been conducted in this domain mainly includes works related to the development of methods and algorithms for addressing the separation problem and the building of datasets for speech separation.

### *2.1. Speech separation methods*

Speech separation algorithms can generally be categorized into audio-only speech separation (AOSS) and audio-visual speech separation (AVSS) methods. Speech separation is traditionally considered a signal-processing problem. Thus, the term audio-only speech separation (AOSS) algorithm means that the proposed methods only exploit the audio signal characteristics to demix the mixed audio. Classical methods such as spectral subtraction (Boll, 1979), computational auditory scene analysis (CASA) (Brownand and Cooke, 1994), and non-negative matrix factorization (NMF) (Paatero and Tapper, 1994; Virtanen, 2007; Cichocki, 2009) have been proven to achieve excellent results. These approaches assume that the speech follows a certain distribution. However, it is difficult in a natural environment to satisfy these assumptions because of the non-stationary behavior of the noise. More recently, several deep learning methods have been proposed, and these approaches have been shown to achieve the best results (Wang and Chen, 2018). Wang and Wang (2013) treated speech separation as a binary classification problem and separated the mixture of the same genders by using deep neural network (DNN) approaches (Wang et al., 2017). Chandna et al. (2017) used a convolutional neural network (CNN) to predict time-frequency masking for speech separation. Fu et al. (2017) proposed a fully convolutional network (FCN) model for raw waveform-based speech enhancement. One difficulty in training the DNN model is that we can not relate separated audio to their corresponding speakers, which is the label permutation problem introduced by Hershey et al. (2016), who proposed a deep clustering method to solve it. Moreover, Yu et al. (2017) proposed permutation invariant training for speaker-independent multi-speaker speech separation. Kolbæk et al. (2017) used recurrent neural networks (RNN) to solve the permutation problem. Luo and Mesgarani (2019) proposed a fully-convolutional time-domain audio separation architecture. However, the AOSS models have difficulty separating similar sounds as the voice of the same gender or the same person at different times (Gabbay et al., 2018).

The audio-visual speech separation (AVSS) algorithms exploit the mixed audio signals with the associated visual embeddings. The natural synchronization between the visual stream and the audio stream in the video is used to guide speech separation. Owens et al. (2018) trained a network to predict whether audio and video streams are aligned in time, which uses the idea of self-supervised learning to adjust the speech separation on or off screen through the learning features extracted by the model. Zhao et al. (2018) proposed PixelPlayer, which utilizes a large number of unlabeled videos to learn to locate the image region that emits sound and separate the sound representing each pixel from the input audio. This method utilizes the natural synchronization of visual and audio modalities to learn models that jointly analyze sound and image without the need for supervisory information. Ephrat et al. (2018) proposed a joint audio-visual model, Looking-to-Listen, which can separate a single voice signal from mixed audio, which includes other speakers and background noise. Visual features are used to "focus" the audio on a specific speaker and improve the quality of speech separation. Gao et al. (2019) proposed a music separation model based on U-Net, which learns object-level sounds from videos containing multiple sound sources. Training deep neural networks to consistently separate audio from similar appearance objects, while accurately reproducing the audio tracks of each sound source object. Li and Qian (2020) used an attention mechanism for multi-modal feature fusion. Unlike previous studies that only used the visual flow of a single target speaker, the paper believes that the visual information of non-target speakers also helps to separate. Therefore, the visual flow of both speakers in mixed audio is input into the model, and then the attention mechanism is used to fuse the visual features of the target speaker and the interfering speaker, achieving simultaneous prediction of two independent speech flows. Chung et al. (2020) proposed a method of replacing the entire video stream with a single facial image and proposed two methods for fusing single image and speech features. One method involves concatenating visual recognition onto each speech embedding frame and verifying its effectiveness in adjusting the separation task through consistency guidance. Another strategy is based on the self-attention mechanism, which provides constraints with different weights for separation. Gao et al. (2021) proposed a multi-modal speech separation model based on U-Net, which simultaneously utilizes lip motion and static facial features as visual modalities to assist in speech separation. By inputting two mixed audio training segments, the model learns the consistency between facial and vocal features from unlabeled videos. Lee et al. (2021) proposed a new cross-modal affinity network that utilizes visual cues to consider relative time information and improve the robustness of speech separation methods to video stream jitter. Montesinos et al. (2022) proposed a low-latency audio-visual two-stage speech separation model. In the first stage, facial motion cues were obtained through lightweight graph convolutional networks, followed by a transformer-based audiovisual speech separation network. In the second stage, pure audio networks were used to enhance the separated audio. Pan et al. (2022) analyzed three different visual sound associations and advocated that time synchronization between speech and its accompanying lip movements is a direct and dominant auditory cue. By simulating human visual attention from top to bottom, they solved the problem of single-channel target speaker extraction. Pan et al. (2022) proposed using co-speech gestures highly related to the conceptual content and prosody of speech to guide speech separation and used two models to achieve speech separation: one is to directly extract the target voice from a mixed sound source, and the other is to first perform speech separation using DPRNN, and then use GSR to explicitly associate the separated speech with the target speaker. This work is somewhat similar to the work in this article, but we use gestures with certain constraints, and it is not necessarily highly related to speech rhythm. Obtaining co-speech gestures, in reality, is not an easy task, so it is difficult to apply this work in practice.

These multi-modal methods have their own strengths, from the use of conditional cues to the fusion methods, which provide good research ideas for later researchers. The research focus of this article is to introduce Chinese sign language modalities to solve the problem of speech separation. The introduction of new modalities also brings new research challenges, among which the most challenging is how to integrate the features of facial, sign language, and speech modalities. The continuous development of audio-visual speech separation technology has led to the development of multi-modal fusion technology, from simple concat to attention mechanisms that are more in line with human brain processing patterns and then to more complex transformer models. This has brought great inspiration to the design of fusion methods in this article, and suitable solutions and comparative experiments have been provided from these three perspectives.

### *2.2. Speech separation datasets*

Various types of datasets can be exploited for solving the speech separation issue, including speech recognition datasets, lip motion recognition datasets, and many audio-visual datasets, including clean audio and corresponding visuals. The common practice is to mix this audio with noises or other clean speeches to get the noisy audio used for the speech separation problem. Garofolo et al. (1993) proposed the TIMIT dataset, which consists of 630 speakers of eight major American English dialects. Patterson et al. (2002) published a speaker-independent corpus CUAVE of 36 speakers saying each digit in a connected and isolated manner. The WSJ0 database proposed by Vincent et al. (2013) was generated from Wall Street Journal news text. The above corpora only contain audio and text modalities. In recent years, audio-visual representation learning has become popular, Cooke et al. (2006) proposed GRID, which is an audio-visual dataset containing 34 speakers, and each speaker says 1000 sentences. The TCD-TIMIT constructed by Harte and Gillen (2015) contains 60 volunteer speakers and is collected using both front-facing and 30-degree cameras. LRS (Chung and Zisserman, 2017a; 2017b; Chung et al., 2017) and VoxCeleb (Nagrani et al., 2020) were recorded by the visual geometry group and contained both high-quality faces and clean audio. Ephrat et al. (2018) constructed a large-scale audio-visual dataset named AVSpeech, which contains thousands of videos spoken by various people, languages, and face poses. Roth et al. (2019) put forward AVA-ActiveSpeaker, which contains about 40K distinct speakers and about 38.5 hours of face tracks.

Table 2. Characteristics of some existing datasets that can be used for speech separation (SS) and sign language recognition (SLR).

| | Datasets | Duration (h) | Speakers | Language | Modalities | Year |
|---|---|---|---|---|---|---|
| SLR | RWTH-PHOENIX-Weather | 12 | 9 | German | Sign Language | 2015 |
| | WLASL | 12 | 100 | English | Sign Language | 2020 |
| SS | TIMIT | 5.3 | 630 | English | Audio; Text | 1993 |
| | CUAVE | 3 | 36 | English | Audio; Text | 2002 |
| | GRID | 27.5 | 33 | English | Audio; Visual; Text | 2006 |
| | CHiME2 WSJ0 | 78 | 101 | English | Audio; Text | 2012 |
| | TCD-TIMIT | 6.3 | 59 | English | Audio; Visual; Text | 2015 |
| | LRS | 800 | 5000 | English | Audio; Visual; Text | 2017 |
| | VoxCeleb | 2000 | 7000 | Diversity | Audio; Visual | 2017 |
| | AVSpeech | 4700 | 1500000 | Diversity | Audio; Visual | 2018 |
| | AVA-ActiveSpeaker | 38.5 | 40000 | Diversity | Audio; Visual | 2019 |
| | CSLNSpeech (Ours) | 10 | 25 | Chinese | Audio; Visual; Chinese Sign Language | - |

Sign language datasets will also be investigated. Koller et al. (2015; 2018) proposed the RWTH-PHOENIX-Weather dataset, which provides RGB videos for full frames and cropped hand gestures in German Sign Language. It is a popular benchmark dataset for continuous sign language recognition, and many types of research (Camgoz et al., 2017; Huang et al., 2018; Pu et al., 2018) have been done on this dataset. Li et al. (2020) introduced a new large-scale Word-Level American Sign Language (WLASL) video dataset containing over 2000 words performed by over 100 signers. The total video duration is about 12 hours. Table 2 summarizes these publicly relevant sign language recognition datasets. As seen from Table 2, the datasets for sign language recognition are generally smaller than those for speech separation. This is because sign language data is more difficult to obtain than audio data, and therefore, it is even more challenging to obtain data with the three modalities.

As far as we know, no speech separation dataset contains both sign language and voice. Therefore, we have built a large multi-modality dataset, CSLNSpeech, which includes voice, face, and Chinese sign language to train our extended speech separation model. We summarize the above-related datasets in Table 2.

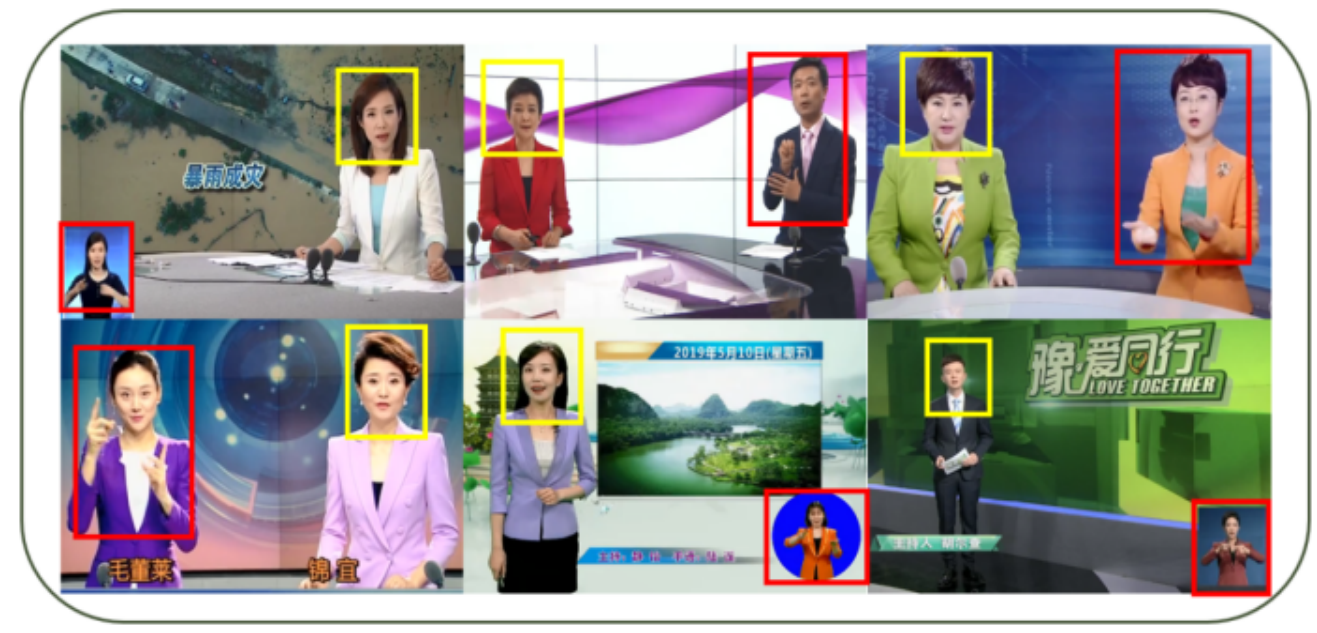

Fig 3 The overview of the source videos where the yellow boxes indicate the visual and the red boxes represent the sign language. The first row is from CCTV, Beijing TV, and Dalian TV, respectively. The second row is from Hangzhou TV, Hangzhou West Lake TV, and Henan TV, respectively.

## 3. CSLNSpeech

In order to explore the role of Chinese sign language in the extended speech separation problem and answer the three questions mentioned above, in this section, we constructed a new large-scale dataset named CSLNSpeech, a multi-modalities dataset, including three modalities of audio, visual, and Chinese sign language. To the best of our knowledge, CSLNSpeech is the first dataset that introduces the sign language modality to the speech separation problem. We think there are at least two reasons why researchers did not introduce sign language to the speech separation problem before (1) The coexistence of audio and sign language modalities from the same person is rare (which can be seen in Fig. 2 (b)); (2) Sign language is a kind of silent language, which is a contrast to audio.

In our research presupposition, sign language only needs the same semantics as audio. It does not need to be from the same person as the audio, significantly reducing the data collection difficulty. Moreover, as mentioned in the introduction, the combination of sign language and audio can help people with hearing loss develop speech and language skills, which plays a practical role in helping those disabled people integrate into society. Therefore, the need to develop methods to extend the speech separation problem makes creating datasets with synchronous audio, visual, and sign language modalities critical. In the following, we will describe the overall process of constructing the CSLNSpeech datasets and give some of its statistics.

### *3.1. Construction of the CSLNSpeech dataset*

Sign language is much more difficult to collect than other modalities like audio and visual. Chinese Sign Language News is one of the most accessible resources for sign language. Therefore, we constructed CSLNSpeech using Chinese Sign Language News and selected the news containing sign language modality from six TV programs in mainland China. We cropped and obtained about 10 hours of TV videos containing the three modalities from about 519 hours of six TV programs. It is not easy to obtain data on three modalities at the same time since the face of the speaker (or announcer) does not always appear in the original video. We need to annotate the time occurrences of the “face-on” and “face-off” of the speaker in the video. Fig. 3 shows the example pictures of six TV programs. As we can see from Fig. 3, audio, visual and sign language can be acquired simultaneously, where the yellow boxes indicate the visual of the speaker and the red boxes represent the sign language of the signer.

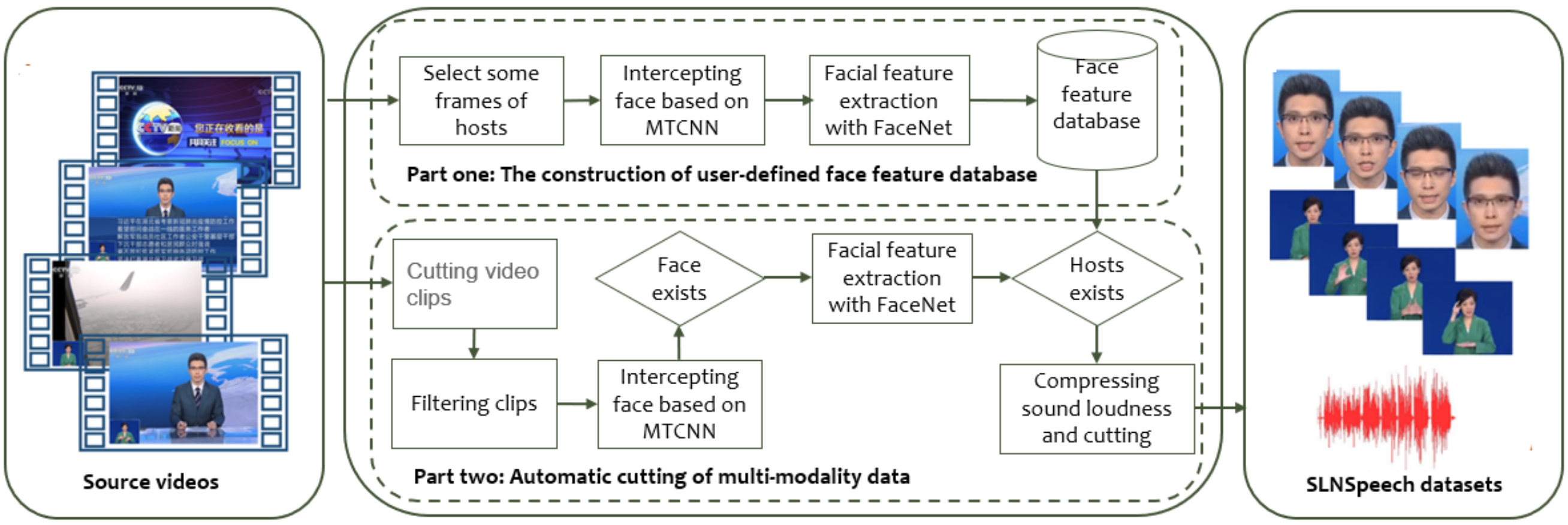

Fig 4 The overall process of constructing CSLNSpeech dataset.

In order to save a large number of labor costs and potential problems caused by wrong annotation, we developed an automatic video clipping toolbox to collect pure audio, visual, and sign language modalities and constructed the CSLNSpeech dataset automatically. The overall process of constructing CSLNSpeech is shown in Fig. 4. The input of the developed video clipping toolbox is shown on the left of Fig. 4, which is about 519-hour source videos from six TV programs that may contain three modalities (audio, visual, and sign language) shown in Fig. 3 or two modalities (audio and sign language) shown in Fig. 2 (c). The output is the separated three modalities of audio, visual, and sign language, which are shown on the right of Fig. 4.

The developed automatic video clipping toolbox is shown in the middle of Fig. 4. It mainly includes two parts:

(1) The construction of the user-defined face feature database

This part will build a user-defined face feature database. Firstly, we manually selected 2,500 video frames containing the face of the 25 announcers, each with 100 discontinuous frames. Then, we used multi-task convolutional neural networks (MTCNN) (Zhang et al., 2016) for face detection to locate the face region in the frames carefully picked. Finally, we used a pre-trained FaceNet (Schroff et al., 2015) to compute and combine every face feature into the user-defined face feature database. The function of the face feature database is to judge whether the cut video clip completely contains the clean data of the face of the announcer, which is the cornerstone of our toolbox.

(2) Automatic cutting of multi-modality data

This part allows our toolbox to automatically clip Chinese sign language news videos that contain audio, visual, and sign language. In the first step, we cut the 519-hour source videos into many clips according to the Perceptual Hash Algorithm (Wen et al., 2010), which can identify adjacent video frames with large pixel changes. Then, since most of the cutting clips acquired from the first step are useless, we picked the cutting clips that meet our requirements, which have the face of the speaker (or announcer) appear throughout the cutting clips. Different from (Hoover et al., 2017), we just detected the first and the last frames of the cutting clips instead of all the frames since the sign language news videos are regular, which means the faces of the speaker are opposite the screen, and the angle of the faces does not rotate too much. For the first and last two frames, we used MTCNN to detect the face and used a pre-trained FaceNet to compute the face feature and compare it with the face feature database when the face is detected. If the faces are in the face database and the faces in the first and last frames are of the same person, we save this cutting clip. Finally, the output is the combination of audio information, for which the loudness of the sounds is normalized to the same level, visual and sign language frames, where the visual and the sign language frames are cut by OpenCV, which is an open-source computer vision library.

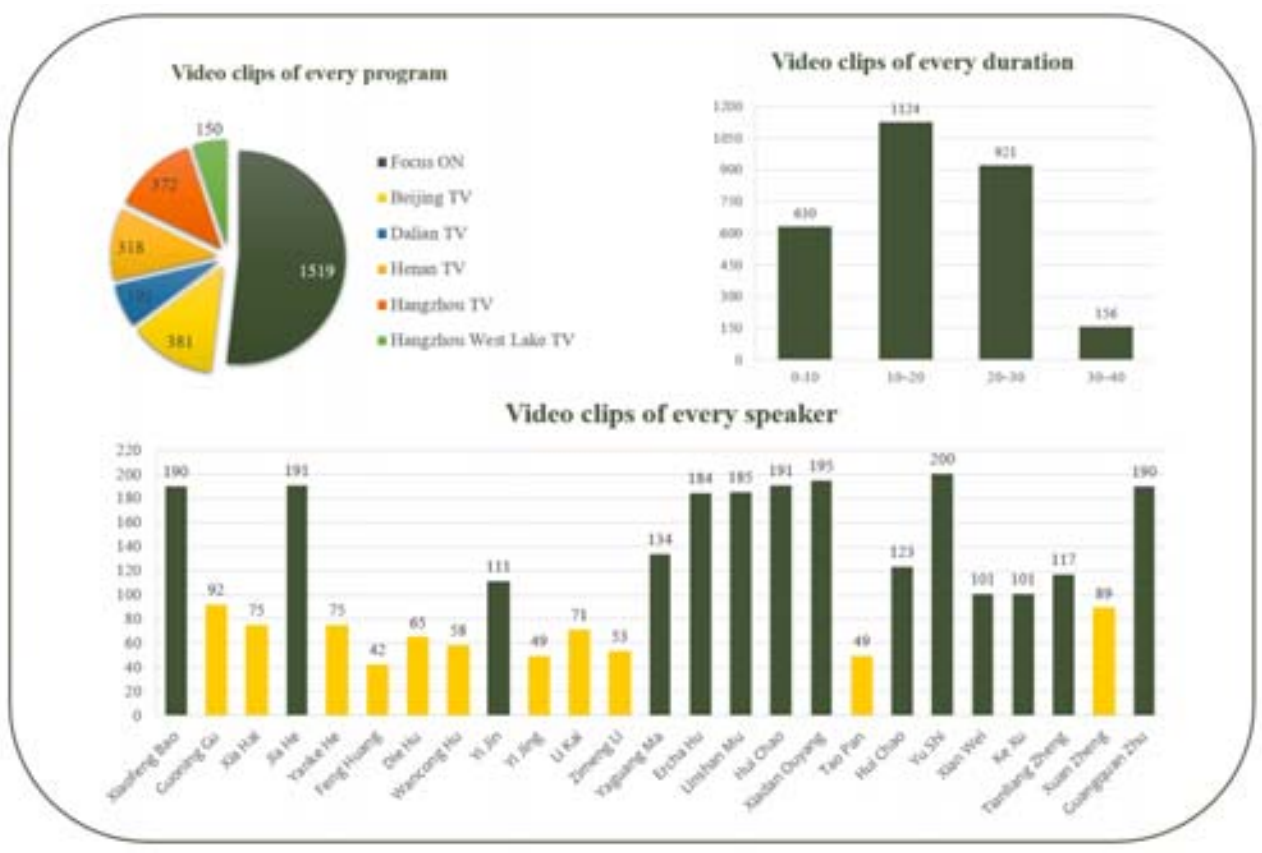

Fig 5 The detailed characteristics of CSLNSpeech dataset in terms of video clip origin, duration, and speakers.

### *3.2. Characteristics of the CSLNSpeech dataset*

The detailed characteristics CSLNSpeech dataset we built are shown in Fig. 5. The repartition of video clips is described through a pie chart, which represents the proportion of the number of segments per TV program. Most videos come from “Focus ON” since the program is well organized so that we can get the source videos conveniently. The duration distribution of video clips is depicted by a histogram where the horizontal axis represents the duration of segments in seconds, and the vertical axis represents the number of segments. The duration of the majority of video clips is between 10 and 30 seconds. In this process, every clip contains 4 or 5 sentences when we randomly mix two clips. The number of video clips per speaker (or announcer) is represented by a bar graph, where the yellow bar is male hosts and the green bar is female hosts. We can see that there are far more video clips for women than for men.

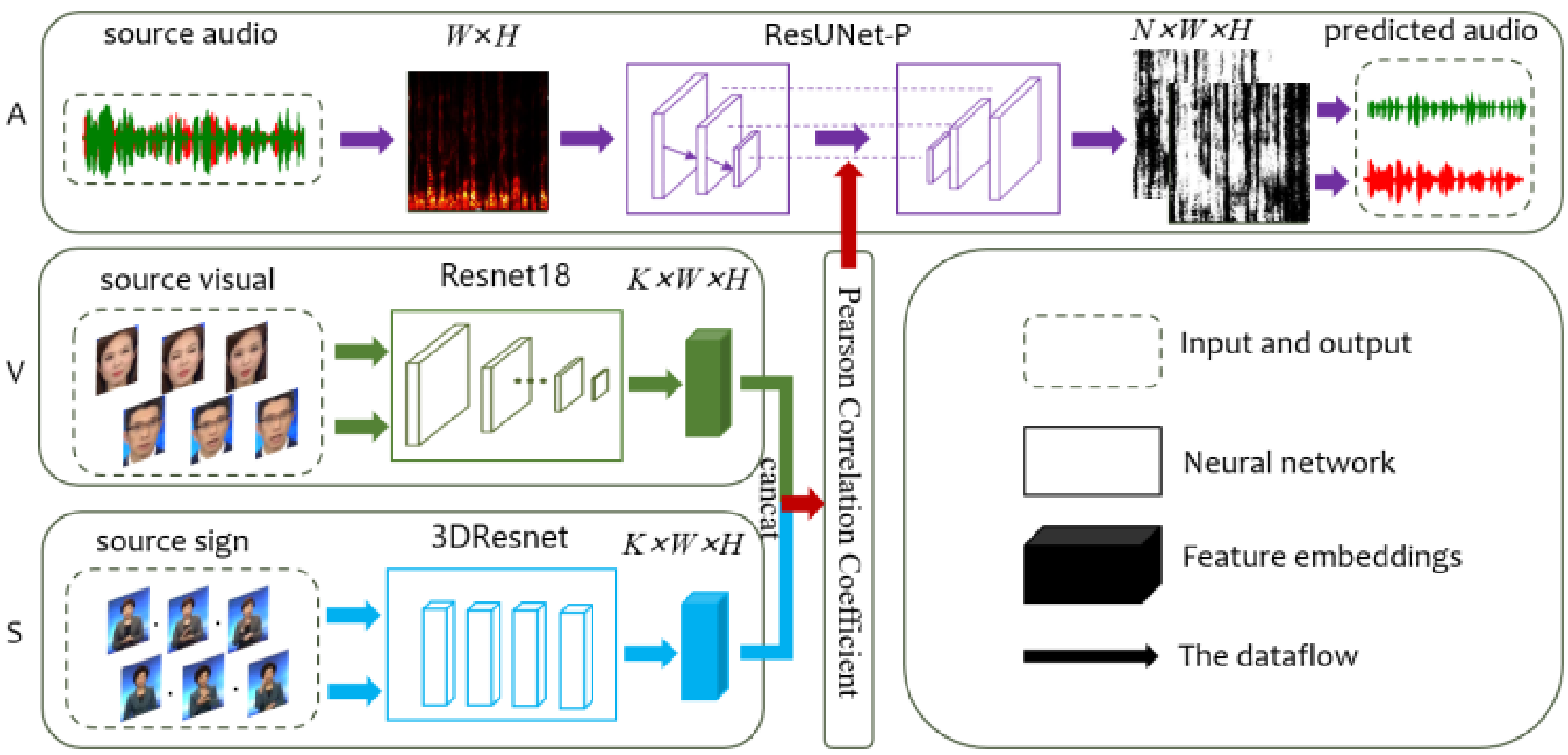

Fig 6 The architecture of the AVSSS network includes three modules: the visual embedding module, the sign language embedding module, and the speech separation module. The sign embedding module and the visual embedding module receive the sign language frames and visual frames and then outputs sign language embedding and visual embedding, respectively. The speech separation module separates mixed source audio according to the sign embedding and the visual embedding. This architecture can operate audio-only network (AO), audio-visual network (AV), audio-sign network (AS), and audio-visual-sign network (AVS). where ⊙ denote the element-wise multiplication.

In total, the CSLNSpeech dataset contains 25 distinct speakers and roughly 10 hours of video segments whose scale is similar to WLASL (Li et al., 2020). The language is mandarin Chinese, and the number of video clips of each speaker is between 42 and 200. Table 2 compares CSLNSpeech with other speech separation datasets and sign language recognition datasets.

## 4. Proposed model

This section describes the speech separation process and the architecture of our AVSSS network. The network architecture includes four components, namely, the visual embedding module, the sign language embedding module, the speech separation module, and the multi-modal feature fusion module, as shown in Fig. 6. The architecture receives three inputs: mixed audio, visual frames, and sign language frames corresponding to audio.

### *4.1. Separating process and training objectives*

We follow the commonly used "mixing and separating" paradigm, mixing clean audio to get mixed audio and taking the corresponding video frames, including face frames and sign language frames, as network input. More formally, given the videos with sign language of speakers, the audio mixture of $n$ speakers can be defined as $\mathrm{x}(t) = \sum_{i=1}^{n} \mathrm{s}_i(t)$, where $\mathrm{s}_i(t)$ are time-discrete signals responsible for each speaker. Our goal is to separate $\mathrm{s}_i(t)$ from $\mathrm{x}(t)$. To perform separation, we predict an ideal binary mask (IBM) (Brownand and Cooke, 1994) $M_i$ for each speaker. First, a Short-Time Fourier Transform (STFT) is applied to the mixture to get its magnitude spectrogram $S_i$. Then the separating process can be defined as:

$$\overline{M}_i = \mathrm{F}(S_i, VF_i, VL_i) \quad (1)$$

where $\overline{M}_i$ denote the predicted ideal binary mask, F denote the whole proposed model, $VF_i$ and $VL_i$ are corresponding face frames and sign language frames in the video, respectively.

The training loss of the model is Binary CrossEntropy Loss (BCELoss) between the predicted mask and the real mask:

$$l = -\frac{1}{n}\sum_{i=1}^{n}[M_i \odot \log(\overline{M}_i) - (1 - M_i) \odot \log(1 - \overline{M}_i)] \quad (2)$$

We then get the element-wise multiplication of the predicted mask and the mixture magnitude spectrogram as the predicted magnitude spectrogram. This utilizes the phase insensitivity of the human ear and generally only predicts the amplitude spectrum of the short-time Fourier transform. The audio waveform is obtained by performing inverse STFT (ISTFT) on the predicted magnitude spectrogram and phase spectrogram from the mixture.

It is worth mentioning that masking is not only calculated using IBM, but also commonly used methods such as IRM. IRM directly characterizes the ratio of pure speech energy to noisy speech energy within each time-frequency unit. In Appendix E, we provide the results of using IBM and IRM. For our model, the actual effects of different masking types are similar. We believe that the type of masking used in our model is not the key to improving performance. However, different prediction targets should be available as options, and selecting targets that match the modeling ability of the model can achieve optimal enhanced performance.

### *4.2. Visual embedding module*

The visual embedding module takes $p$ facial frames, which are randomly selected in the same time interval from the video, as the network input. The visual embedding module consists of a ResNet18 (He et al., 2016) removed the last fully-connected layer and a temporal pooling layer with sigmoid activation. It extracts a facial feature representation of dimension $k/2 \times w \times h$, where $k$ is the number of channels. We try to replace ResNet18 with ResNet50 to verify whether the deeper neural network will improve speech separation better, and we find that this change does not have an apparent impact on speech separation. The relevant experiment results are shown in Appendix F.

### *4.3. Sign language embedding module*

The sign language embedding module takes $p$ sign language frames as input, which correspond to the facial frames in time. For the sign language embedding module, we use 3DResNet (Pu et al., 2018), which is proven to be able to learn spatio-temporal features to extract hand motion features, to extract sign language feature representation of dimension $k/2 \times w \times h$, where $k$ represents the number of channels.

### *4.4. Speech separation module*

For the speech separation module, we use ResUNet, a combination of ResNet (He et al., 2016) and UNet (Rahmani et al., 2018), that is, UNet with residual connection in every encoder block. The ResUNet consists of an encoder and a decoder network, with jump connections between the encoder and decoder. The encoder takes the mixture spectrogram of dimension $1 \times f \times t$, where $f$ represents frequency resolution and $t$ represents time resolution of the spectrogram, as input and output an audio feature map of dimension $k \times w \times h$. Then the multi-modality fusion module fuses the audio feature map, the visual embedding, and the sign language embedding by calculating an attention matrix base on the Pearson correlation coefficient (PCC) (detailed description in Section 4.5), which is added to the audio feature map to get the fusion feature. The fusion feature is passed through the ResUNet decoder to restore the original resolution. The final output is the predicted mask of dimension $1 \times f \times t$.

### *4.5. Multimodal feature fusion module*

We propose to use a new multi-modality fusion method, the Pearson correlation coefficient (PCC) based attention fusion method, which is implemented by calculating the PCC coefficient of facial, sign language and audio features, measuring the correlation between visual and audio. We first concatenate the facial embedding and sign language embedding along the channel dimension to generate a visual feature of dimension $k \times w \times h$. Then we calculate the PCC coefficient of the visual feature and the audio feature from the ResUNet encoder (as described in Section 4.4):

$$\mathrm{PCC}(A, V) = \frac{\mathrm{E}[(A-\mu_A)\odot(V-\mu_V)]}{\sigma_A-\sigma_V} \quad (3)$$

where $A$ is the audio feature, $V$ is the visual feature, E indicates expectation, $\mu_*$ means the mean value of $*$, and $\sigma_*$ means the standard deviation of $*$, where $*$ refers to $A$ and $V$.

The $\mathrm{PCC}(A, \mathrm{V})$, we regard as a spatial attention matrix, is a matrix with values between -1 and 1, in which positive values indicate the positive correlation between *A* and *V*. Thus, the ReLU activation is applied to the attention matrix to keep the positive correlation. Each value in the attention matrix is calculated through the channel vector at that location, which can effectively represent the correlation between pixel features and time-frequency features in the potential space. We get the fused feature by adding the attention matrix to the audio features.

To verify the advantages of the PCC-based attention fusion method, we also tried simple concatenation (concat) and Transformer-based fusion methods. The former simply combines the features of the different modalities along the channel dimension. The latter adopts a method similar to Vision Transformer (Dosovitskiy et al., 2020). Firstly, the visual and audio features are concat along the channel dimension and then partitioned into patches in the spatial dimension. This process is achieved through a linear projection layer, where the patches are sequentially unfolded to obtain a sequence of features, which are input into a stacked transformer encoder layer for deep fusion. Through the transformer, correlations between the patch features can be calculated.

## 5. Experiments and results

### *5.1. Datasets*

The CSLNSpeech dataset includes audio, visual, and Chinese sign language modalities. Therefore, we use the CSLNSpeech dataset to evaluate the performance of our proposed AVSSS architecture and mainly study the role of sign language modality. The GRID dataset (Cooke et al., 2006) and the TCD-TIMIT dataset (Harte et al., 2015) are two open-source datasets containing audio and visual modalities. It is worth noting that our AVSSS architecture is compatible and can perform speech separation without sign language modality. Therefore, we conduct audio-visual speech separation experiments on the GRID and TCD-TIMIT datasets to prove our proposed model's effectiveness without sign language modality. We extracted about 3 seconds of audio clips from each speaker and randomly combined clips from different speakers. Ultimately, we generated 13,200 audios, composite mixtures for our model (4400 male-male, 4400 female-female, and 4400 male-female pairs). For each experiment, the resulting dataset was split into disjoint 12000 synthetic audio mixture sets for training and 1200 synthetic audio mixture sets for testing.

### *5.2. Implementation details*

For the visual embedding module and sign language embedding module, the corresponding frame sizes are $3 \times 3 \times 224 \times 224$ and $3 \times 3 \times 140 \times 140$, respectively, where the first 3 represent three frames and the second 3 represent three channels. The output size of the sign language embedding module and the visual embedding module are both $k/2 \times w \times h$, where $k$ is set to 512 in the experiments.

For the speech separation module, we resampled all audio to 16kHz, and STFT, with 25ms duration and 40% overlapping sliding window at 16kHz sampling rate, is used to obtain a spectrum map of dimension $512 \times 320$. The size of the input spectrogram is $1 \times 512 \times 320$. Then, the spectrogram is fed into a ResUNet with five convolutional layers and five deconvolutional layers. ResUNet outputs a binary mask after sigmoid activation and $n$ binary masks of size $1 \times 512 \times 320$ for $n$ inputs of the mixed audio stream.

We used the SGD optimizer with momentum 0.9 and weight decay 1e-4 to train our model for 150 epochs, with the learning rate initially set to 1e-1 and reduced by 0.1 times at the 40th and 80th epochs, respectively.

Table E.1. AVS experiments with different fusion methods on SLNSpeech dataset

| | Fusion Type | SI-SDR | SDR | SIR | | SAR | PESQ | STOI | FLOPs(G) | Params(M) |
|---|---|---|---|---|---|---|---|---|---|---|
| AVS | Concat | 8.96 | 8.97 | 15.81 | | 10.28 | 2.43 | 0.87 | 140.83 | 60.81 |
| AVS | Transformer | 9.14 | 9.14 | 16.09 | | 10.40 | 2.46 | 0.88 | 192.91 | 142.28 |
| AVS | PCC | 9.12 | 9.12 | 16.03 | | 10.40 | 2.46 | 0.88 | 140.83 | 60.81 |

The BCELoss is used to predict the ideal binary mask of the audio spectrogram. The batch size of our experiments is 5.

### *5.3. Results and analysis*

Speech separation results are evaluated by using a series of standard metrics, including signal distortion ratio (SDR), signal interference ratio (SIR), and signal artifact ratio (SAR) from the mir_eval library (Raffel et al., 2014). We have also included two speech-specific metrics: Perceived Speech Quality Assessment (PESQ) (Rix et al., 2001), which measures the overall perceived quality of separated speech, and short-term objective intelligibility (STOI), which is a metric related to signal intelligibility (Taal et al., 2011). The specific definitions can be found in Appendix G.

#### *5.3.1. Experiments on different fusion methods*

We compare three different fusion methods mentioned in the paper on the AVS method on the SLNSpeech dataset. The results are shown in Table 3. In addition to the performance indicators of speech separation, we also introduce Floating point operations (FLOPS) and parameter quantities for further comparison. FLOPs can be used to measure the computational complexity of models. The parameter quantity can be used to measure the size of the model. Considering that the result may be very large, we use G as the unit of FLOPs to indicate how many times the model calculates billions per second and use M as the range of Params to indicate how many millions of parameters the model has. We expect that the model can be as simple and fast as possible while ensuring its performance of the model.

It can be seen that, in terms of speech separation performance, the PCC fusion method is better than the concat method and slightly lower than the transformer-based method. However, the FLOPS and parameters of the PCC fusion method are much lower than the transformer-based method, which is the same as the concat method. This indicates that the PCC method can effectively calculate the correct multimodal feature correlation without the need for additional parameters and computational complexity. Based on this, we also use the PCC fusion method in subsequent modal ablation experiments.

#### *5.3.2. Experiments on the CSLNSpeech dataset*

We verify the effect of the multi-modality speech separation system with sign language through the following experiments on the CSLNSpeech dataset:

- uPIT: A baseline method for pure audio speech separation in the time and frequency domain.
- AO: The experiment uses UNet, which has the same architecture as our speech separation module, to do the audio-only separation.
- AS: The experiment in which our model takes the audio mixture and sign language frames as input to separate the speech.
- AV: The experiment in which our model takes mixed audio and face frames as input to separate the speech.
- AVS: The experiment in which our model takes mixed audio, face frames, and sign language frames as input to separate the speech.

Besides the above several modality ablation experiments, we use fake faces to verify whether sign language can be attacked in visual modality to ensure the robustness of the separation system:

- 50% fake faces (part): This means that we include 50% of the fake faces in the dataset, and our model will use one fake face frame when encountering fake faces, simulating that part of the video used in speech separation is attacked by deepfake. We conduct AV and AVS experiments under the data condition.
- 50% fake faces (all): This simulates that the whole video used in speech separation is attacked by deepfake. AV and AVS experiments are conducted.

We present other percentages of false face data in Appendix H. In this paper, we select the median value of 50% to analyze.
modality fuses on speech separation compared to an audio-only baseline. We also present the results of speech separation under different degrees of deepfake attacks, including partially attacked, that is, one frame of the visual frame is replaced with a fake face, and completely attacked, that is, all the visual frames are replaced with a fake face.

The second line of Table 4 reports the results of the best method of frequency domain audio-only speech separation; the third line reports the results of only using audio input; the 4th line reports the results of using audio and sign language as input. It can be seen that the AS method is better than the AO method as well as the baseline method (uPIT). Therefore, we conclude that sign language can be used to supervise speech separation. For the model learning, hand movement information provided by sign language modality proves paramount in ensuring critical alignment of the speaker's voice, which has supervisory implications for speech separation.

The 5th line reports the results of using audio and visual as input, which is better than AO and AS. The 6th line reports the results of using audio, visual, and sign language as input, which is a little better than AV. Therefore, we speculate that sign language is a weak supervision signal compared with the face as a strong supervision signal in audio-visual learning. These results verify the correctness and rationality of the proposed speech separation network method. It shows that the network has learned the relatively strong correlation between audio and vision and the relatively weak correlation between audio and sign language.

The 8th and the 9th lines show the AV separation results and AVS separation results of part fake face attacks, respectively. It can be seen that the performance of AV decreases slightly, while AVS shows good robustness. The 11th and 12th lines show the results of the AV and AVS experiments that are completely attacked. Compared with the experiments that were not attacked by fake faces, the separation performance indicators of the AV and AVS methods have decreased, but AVS shows better robustness because the SDR of AV decreases by 1.33db, while the SDR of AVS only decreases by 0.56db. It can be concluded that sign language can be used as a critical auxiliary supervision method for speech separation when the correct visual modality is missing.

Finally, we draw the following conclusions on the question of sign language modality proposed in the introduction:

Table 4 shows the effect of the proposed model on the CSLNSpeech dataset, where we evaluate the performance impact of different multi-

Table 4. The comparison of different modalities in speech separation using CSLNSpeech dataset. The table can be divided into three parts: the common comparison, no fake face and 50% fake faces. AO: audio-only; AV: audio-visual; AVS: audio-visual-sign.

| | SI-SDR | SDR | SIR | SAR | PESQ | STOI |
|---|---|---|---|---|---|---|
| uPIT (Yu et al., 2017) | 8.46 | 8.48 | 15.53 | 9.83 | 2.28 | 0.86 |
| AO (Rahmani et al., 2018) | 6.21 | 6.24 | 11.80 | 7.62 | 1.91 | 0.76 |
| AS | 8.96 | 8.97 | 15.76 | 10.29 | 2.43 | 0.87 |
| AV | 9.10 | 9.11 | 16.04 | 10.39 | 2.45 | 0.87 |
| AVS | 9.12 | 9.12 | 16.05 | 10.40 | 2.46 | 0.88 |
| 50% fake faces (part) | | | | | | |
| AV | 9.06 | 9.08 | 15.96 | 10.36 | 2.44 | 0.87 |
| AVS | 9.12 | 9.12 | 16.06 | 10.39 | 2.45 | 0.87 |
| 50% fake faces (all) | | | | | | |
| AV | 7.77 | 7.79 | 14.13 | 9.61 | 2.21 | 0.84 |
| AVS | 8.55 | 8.56 | 15.26 | 10.07 | 2.37 | 0.86 |

(1) Can sign language supervise speech separation like visual information does?

The answer is yes, as the result of ASSS is better than that of AOSS, so sign language can supervise speech separation and improve performance. Meanwhile, the result of AVSS is better than that of ASSS, so sign language is a kind of weak supervision compared with face modal in speech separation, which is consistent with our expectations.

(1) Is the performance of AVSSS better than AVSS?

The answer is yes, because in the CSLNSpeech dataset we built, sign language and audio are synchronized to a certain extent, and AVSSS performs slightly better than AVSS.

(3) Can the weakly supervised sign language improve the robustness of the AVSSS?

The answer is yes because AVSSS is better than AVSS when being attacked by deepfake technology. When visual modality is absent or attacked, sign language can be used as critical auxiliary supervision means in speech separation.

#### *5.3.3. Comparison with audio-visual speech separation baselines*

To verify the effectiveness of our model, we conducted experiments on two other open-source audio-visual datasets, that is, GRID and TCD-TIMIT, which are introduced in related work. Since there is no sign language modality, we tested our AV model and compared it with the latest audio-visual speech separation methods. The results are shown in Table 5. It can be seen that our AV model is superior to other baseline methods in both datasets, which indicates the advantage of our model architecture.

## 6. Discussion

What might be confusing about the proposed method is the conflict between the silence characteristic of sign language and audio. The first thing to be clear is that sign language-assisted speech separation can help those hearing loss people to learn speech, for instance, by removing the interfering sound in the pronunciation practice video of the supporting sign language for the hearing impaired, and those hearing-impaired people with language ability trend to use sign language simultaneously while speaking. If a robot or intelligent agent can process multiple modal information, including sign language, it can better help these disabled people.

Table 5. The comparison results of speech separation on GRID and TCD-TIMIT datasets.

| Datasets | Methods | SDR | Params | Flops |
|---|---|---|---|---|
| GRID | Morrone et al. (2019) | 8.05 | - | - |
| | **Ours (AV)** | **8.88** | 45.38 | 134.34 |
| TCD-TIMIT | Gabbay et al. (2018) | 0.4 | - | - |
| | Ephrat et al. (2018) | 4.1 | 24.35 | 0.05 |
| | Morrone et al. (2019) | 10.55 | - | - |
| | Gao et al. (2021) | 10.9 | 76.31 | 41.86 |
| | **Ours (AV)** | **12.16** | 45.38 | 134.34 |

Moreover, taking sign language translation technology as a comparison, the goal of sign language translation is to translate sign language into corresponding semantic words, which is a more direct and accurate way to help people with hearing disabilities communicate. However, the difficulty of sign language translation is that sign language is not translated verbatim according to the language but needs to be refined and word order adjusted according to the overall meaning of the sentence. It faces incredibly complex technical problems regarding information conciseness, low latency, and refinement. Creating a sign language translation dataset is also very difficult and requires much manpower. In our research, we do not rely on the semantics of sign language but on the synchronization of sign language and voice in time (ensuring semantic synchronization). Our focus is on the joint learning and fusion of multiple modalities. We also propose an automated method for building multi-modality datasets. We hope to contribute to the research of more general models in the multi-modality community.

## 7. Conclusion

In this paper, we consider an extended speech separation problem which refers, in particular, to sign language assisted speech separation and the separation of the audio with the help of different modalities by extracting visual and sign language features. To solve this problem, we created a new large-scale dataset, CSLNSpeech, that gathers audio, visual, and Chinese sign language modalities collected from the sign language news of TV programs. We propose a general AVSSS network to solve the extended speech separation problem. The evaluation results show that sign language plays a weak role in speech separation compared with face, and it works well when visual modality is missing or attacked. Moreover, our model has achieved the best results on GRID and TCD-TIMIT datasets. The positive results of sign language assisted speech separation also open a door, which means that more modalities can be considered when effectively solving different problems.

## CRediT authorship contribution statement

**Jiasong Wu:** Conceptualization, Writing- Reviewing and Editing. **Xuan Li:** Writing- Original draft preparation, Software, Visualization. **Taotao Li**: Methodology, Software. **Fanman Meng:** Software, Visualization. **Youyong Kong:** Data Curation, Validation. **Guanyu Yang:** Validation, Project administration. **Lotfi Senhadji:** Formal analysis, Writing- Reviewing and Editing. **Huazhong Shu:** Supervision, Resources.

## Declaration of competing interest

The authors declare that they have no known competing financial interests or personal relationships that could have appeared to influence the work reported in this paper.

## Acknowledgements

This work was supported in part by the National Key Research and Development Program of China (Nos. 2021ZD0113204, 2022YFE0116700), in part by the National Natural Science Foundation of China under Grants 62171125 and 61876037, in part by the Industry-University-Research cooperation project of Jiangsu Province under grant BY2022564, and in part by INSERM under the Grant the calls IAL and IRP. The authors would like to thank the Big Data Computing Center of Southeast University for providing the facility support on the numerical calculations in this paper. The authors would like to thank the 25 hosts, 23 signers and the 6 TV stations mentioned above for their source videos (see the supplement document). We are also thankful to Kensho Hara for providing his Python code constructing the 3DResNet and Hang Zhao for his open source codes of sound of pixels.